\documentclass[aps,prl,notitlepage,twocolumn]{revtex4}

\usepackage{amssymb}
\usepackage{amsmath}
\usepackage{graphicx}
\usepackage{bm}
\usepackage{mdframed}
\usepackage{color}
\usepackage[mathscr]{euscript}
\usepackage{verbatim}
\usepackage{stmaryrd}
\usepackage[colorlinks=true]{hyperref}

\newcommand{\ud}{\,\mathrm{d}}

\begin{document}

\bibliographystyle{prsty}
\author{Ricardo Zarzuela, Se Kwon Kim, and Yaroslav Tserkovnyak}
\affiliation{Department of Physics and Astronomy, University of California, Los Angeles, California 90095, USA}

\title{Stabilization of the skyrmion crystal phase in thin-film antiferromagnets}

\begin{abstract}
We investigate the formation and stability of the skyrmion crystal phase in antiferromagnetic thin films subjected to fieldlike torques such as, e.g., those induced by an electric current in CuMnAs and Mn$_{2}$Au via the inverse spin-galvanic effect. We show that the skyrmion lattice represents the ground state of the antiferromagnet in a substantial area of the phase diagram, parametrized by the staggered field and the (effective) uniaxial anisotropy constant. 
Skyrmion motion can be driven in the crystal phase by the spin transfer effect. In the metallic scenario, itinerant electrons experience an emergent SU$(2)$-electromagnetic field associated with the (N\'{e}el) skyrmion background, leading to a topological spin-Hall response. Experimental signatures of the skyrmion crystal phase and readout schemes based on topological transport are discussed.
\end{abstract}

\maketitle 

{\it Introduction.}|Skyrmions are the quintessence of spatially localized solitons in quasi-two-dimensional (quasi-2D) spin systems \cite{Belavin-JETP1975}. These magnetic textures exhibit a particlelike behavior \cite{Pollath-PRL2017,Lin-PRB2013}, carry topological charge and are protected against structural distortions and moderate external perturbations \cite{Rohart-PRB2016}. Skyrmions arise in magnetic systems with broken inversion symmetry and spin-orbit coupling \cite{Bogdanov-PRB2002,Banerjee-PRX2014}, and have been observed in a plethora of ferromagnetic materials \cite{Mulbauer-Sci2009,Yu-Nat2010,Seki-Sci2012,Jiang-Sci2015}. The recent years have witnessed a growing interest in these topological excitations due to their potential usage as building blocks for logic devices and information storage \cite{Fert-NatNano2013,Zhou-NatComm2014,Zhang-SciRep2015,Chen-2017,Fert-NatRevMat2017}, controllable nucleation/annihilation by local spin-polarized current injection \cite{Sampaio-NatNano2013}, low current threshold for depinning \cite{Nagaosa-NatNano2013} and unconventional transport properties such as the skyrmion Hall effect \cite{Jiang-NatPhys2017,Litzius-NatPhys2017}. Furthermore, along with the Abrikosov vortex lattice in type-II superconductors \cite{Abrikosov}, the skyrmion crystal phase (SkX) stands out as almost the only well-understood example of soliton crystal, therefore illustrating the crystal order beyond the usual atomic/molecular paradigm.

Antiferromagnets provide an optimal platform to exploit skyrmions since they display ultrafast spin dynamics (with frequencies lying in the THz range) and produce minimal stray fields. This scenario has been intensively investigated in recent years, yielding qualitatively similar results to those found for ferromagnets on the topological robustness and the creation/annihilation (by spin currents) of these solitons \cite{Keesman-PRB2016,Zhang-SciRep2016}. On the contrary, intrinsically different (current-/thermally driven) dynamics are obtained since the gyrotropic response of skyrmions is suppressed in antiferromagnetic (AFM) materials \cite{Kim-PRB2015,Baker-PRL2016,Velkov-NJP2016,Jin-APL2016}. Remarkably, this suppression allows for significantly larger values of the (terminal) skyrmion velocity as compared to the ferromagnetic case \cite{Baker-PRL2016,Velkov-NJP2016}, which represents an attractive feature from the standpoint of technological applications. Nevertheless, no AFM skyrmion crystal phase has been observed yet. The underlying main reason is that the staggered order parameter couples weakly to electromagnetic fields and, therefore, spin textures in the AFM phase are generally not easy to drive or read out. 

Hitherto, it is largely unknown whether the SkX phase can be stabilized in thin-film antiferromagnets and, if so, which of its macroscopic signatures are accessible experimentally. Further insight into these questions is thus vital to boost progress in the field of skyrmion spintronics. In this Letter, we explore different possibilities to utilize fieldlike torques for the stabilization of the SkX phase in quasi-2D AFM films. We show that the fieldlike torques emulate a Zeeman coupling for the N\'{e}el order, $\bm{l}$, at the level of energetics, $\mathcal{F}_{\textrm{Z}}[\bm{l}]=$ $-\bm{l}\cdot\mathcal{B}_{\textrm{stag}}$, where $\mathcal{B}_{\textrm{stag}}$ denotes the staggered field. Generically, there are two possible ways of inducing this staggered field: First, by endowing an effective ferromagnetism in the film, so that $\mathcal{B}_{\textrm{stag}}$ corresponds to an equilibrium magnetic field. One example is offered by Cr$_{2}$O$_{3}$ thin films where the magnetoelectric effect gives rise to a boundary ferromagnetism \cite{Binek-NatMat2010}. 
Second, by preserving the AFM nature of the system and breaking (structural) symmetries that allow the onset of the staggered field via nonequilibrium (electrical) effects. The latter is the case of Mn$_{2}$Au and CuMnAs, where $\mathcal{B}_{\textrm{stag}}$ is induced by the inverse spin-galvanic (Edelstein) effect \cite{Zeleny-PRL2014,Wadley-Sci2016}. One of our main results is the free-energy density
\begin{equation}
\label{eq1}
\mathcal{F}_{\textrm{eff}}[\bm{l}]=\frac{A}{2}\sum_{\mu=x,y}(\partial_{\mu}\bm{l})^{2}+\frac{\mathcal{K}}{2} l_{z}^{2}+\mathcal{F}_{\textrm{DM}}[\bm{l}]-l_{z}\mathscr{B}_{\textrm{stag}},
\end{equation}
which describes AFM films with broken inversion symmetry along the normal to the basal plane ($z$ axis), and with in-plane isotropy in spin space. The terms on the right-hand side represent, from left to right, the exchange interaction, the uniaxial (along the normal to the film) effective anisotropy, the Dzyaloshinskii-Moriya (DM) interaction \cite{Dzyaloshinskii-JETP1957,Moriya-PR1960,Dzyaloshinskii-JETP1964} and the Zeeman-like term, with the staggered field lying along the $z$ axis. 

\begin{figure*}[t]
\begin{center}
\includegraphics[width=\textwidth]{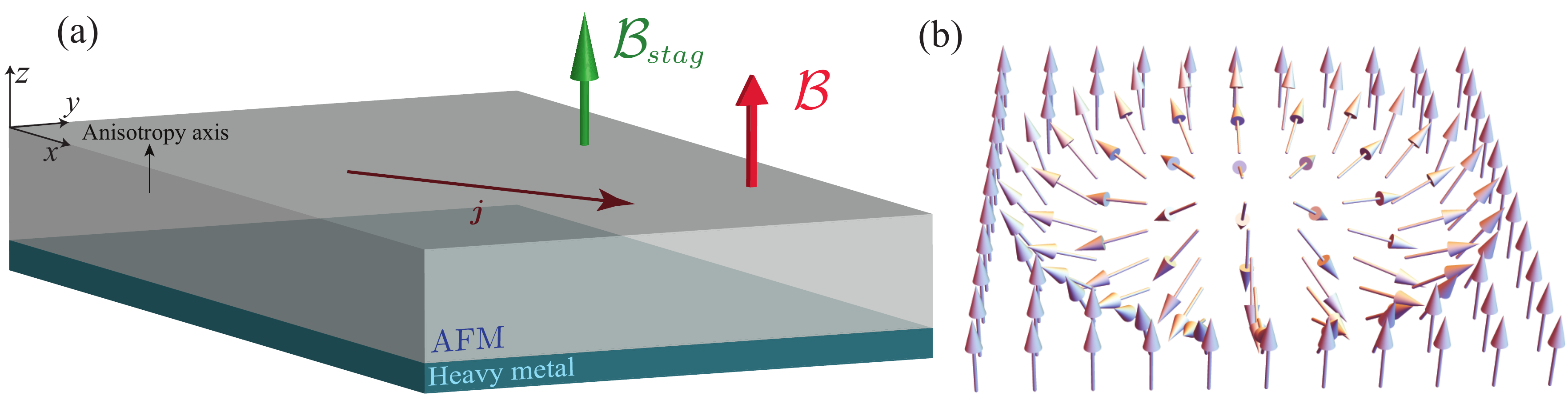}
\caption{(a) AFM thin film deposited on a heavy-metal substrate, whose MCA easy axis lies along the normal to the film ($z$ axis). (b) Spatial dependence of the staggered order parameter for a single N\'{e}el skyrmion.}
\label{Fig1}
\end{center}
\end{figure*}

This last term, as we will elucidate below, becomes crucial for the stabilization of the AFM SkX phase (as in the ferromagnetic counterpart \cite{Banerjee-PRX2014}). In particular, we demonstrate that the skyrmion lattice is the ground state in a substantial area of the phase space $\mathcal{M}(\mathcal{K},\mathscr{B}_{\textrm{stag}})$, possessing thus a large degree of tunability. We also find signatures of this skyrmion phase in the spin-Hall response of conduction electrons. Finally, we propose the readout of the SkX phase in AFM insulators via nonlocal transport measurements \cite{Ochoa-PRBRC2017}.       

{\it Effective theory.}|It can be readily seen that any fieldlike torque $\bm{\tau}_{\textrm{FL}}=\bm{l}\times\mathcal{B}_{\textrm{stag}}$ driving the spin dynamics of the antiferromagnet engenders a Zeeman-like term for the N\'{e}el order at the level of energetics \cite{SM}. Note that these reactive torques arise when sublattice symmetry is broken. Within the exchange approximation, the minimal model for the free energy of the AFM thin film is given by Eq. \eqref{eq1}, which corresponds to the configuration of both external magnetic and staggered fields parallel to the $z$ axis \cite{SM}. Film thickness is taken to be less than the AFM exchange length, so that we can safely assume the uniformity of the N\'{e}el order along the $z$ axis. The (in-plane isotropic) exchange term is described by the stiffness constant $A$ and the effective anisotropy has two contributions, one given by the magnetocrystalline anisotropy (MCA) of the film and the other rooted in the weak coupling of the N\'{e}el order to the external magnetic field $\mathcal{B}$. The effective anisotropy constant for the uniaxial case reads as $\mathcal{K}(\mathscr{B})=K+\chi\mathscr{B}^{2}$, where $\chi$ and $K$ denote the (transverse) spin susceptibility and the on-site uniaxial MCA constant, respectively. For $K<0$ (easy-axis antiferromagnet), our minimal model undergoes the spin-flop transition at the magnetic field $\mathscr{B}_{F}=\sqrt{|K|/\chi}$, where $\mathcal{K}$ flips its sign.

We also include an interfacial DM interaction in our description of the AFM film, which is provided by an adjacent heavy metal breaking reflection symmetry along the $z$ axis, see Fig. \ref{Fig1}(a). It is given by the Lifshitz invariant $\mathcal{F}_{\textrm{DM}}[\bm{l}]=D(l_{z}\nabla\cdot\bm{l}-\bm{l}\cdot\nabla l_{z})$, where $D$ denotes the Dzyaloshinskii coupling constant. Bulk DM terms of the form $D_{\textrm{b}}\,\bm{l}\cdot(\nabla\times\bm{l})$ are also allowed when (global) centrosymmetry is broken. Note that a global spin rotation (described by the Euler axis-angle $\vartheta_{\textrm{DM}}\hat{e}_{z}$, with $\tan\vartheta_{\textrm{DM}}=D_{\textrm{b}}/D_{\textrm{int}}$) maps the total DM energy onto an effective interfacial term, while leaving the other contributions to the free-energy functional invariant \cite{Kim-2018}. As a result, in the rotated spin frame of reference, the AFM film is described by Eq. \eqref{eq1} with an effective Dzyaloshinskii constant $D=\sqrt{D_{\textrm{b}}^{2}+D_{\textrm{int}}^{2}}$; this model, as we show below, stabilizes skyrmions of the N\'{e}el type. In the original (unrotated) spin frame of reference, however, the stable topological textures resemble a hybrid of N\'{e}el and Bloch skyrmions, as can be readily seen from applying the inverse spin rotation to the ansatz for the former. 

Single N\'{e}el skyrmions, depicted in Fig. \ref{Fig1}(b), are metastable solutions of the exchange energy functional, since the latter corresponds to the two-dimensional $O(3)$ nonlinear $\sigma$-model for the N\'{e}el order. These magnetic solitons are classified by the Pontryagin index (so-called topological charge) defined per
\begin{equation}
\label{eq2}
\mathcal{Q}_{\textrm{sky}}=\int_{\mathbf{R}^{2}}\hspace{-0.1cm}\ud^{2}\bm{r}\hspace{0.1cm}\rho_{\textrm{sky}},\hspace{0.4cm}\rho_{\textrm{sky}}=-\frac{1}{4\pi}\bm{l}\cdot\left(\partial_{x}\bm{l}\times\partial_{y}\bm{l}\right),
\end{equation}
which is an (integer) invariant providing a measure of the wrapping of the AFM order around the unit sphere. The exchange energy $\mathcal{F}_{\textrm{sky}}=4\pi A|\mathcal{Q}_{\textrm{sky}}|$ of skyrmions is independent of their center and size, a feature rooted in the invariance of this $O(3)$ nonlinear $\sigma$-model under spatial translations and scaling transformations. Collapse of these solitons into atomic-size defects is prevented by the presence of the DM interaction and the uniaxial anisotropy, which introduce a characteristic length scale below which spatial fluctuations of the texture (in particular, shrinking) are energetically penalized: Minimization (at zero fields) of Eq. \eqref{eq1} with account of a hard cut-off variational ansatz for the rigid skyrmion and the constraint $|\mathcal{Q}_{\textrm{sky}}|=1$ for the topological charge yields the value $R_{\star}\simeq2\pi D/K$ for its size \cite{Ochoa-PRB2016}. 

\begin{figure}[t]
\begin{center}
\includegraphics[width=8.8cm]{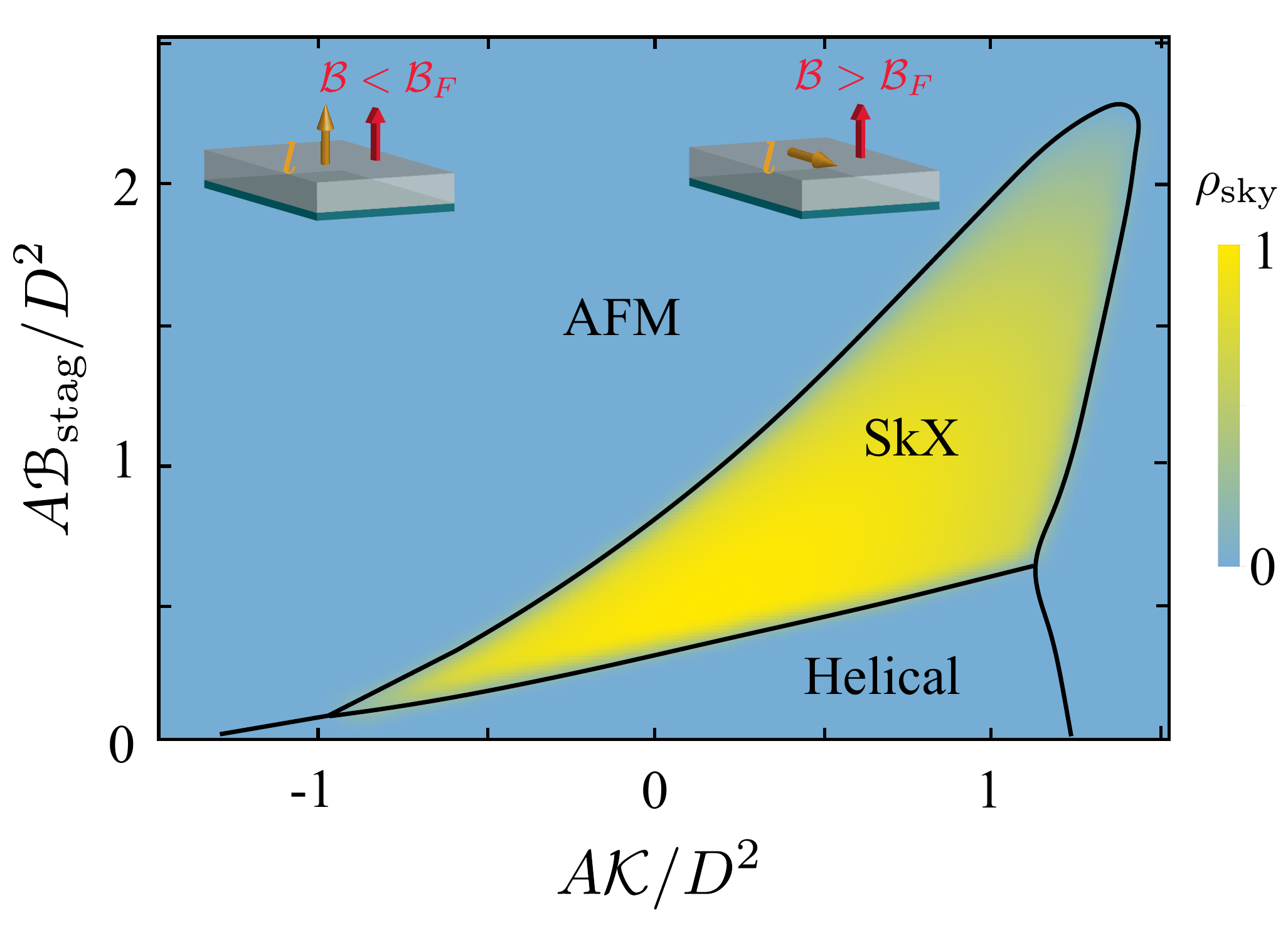}
\caption{Zero-temperature phase diagram in the anisotropy-staggered field phase space $\mathcal{M}(\mathcal{K},\mathscr{B}_{\textrm{stag}})$. It is obtained from a one-dimensional circular-cell minimization of the free energy \eqref{eq1}, with ensuing uniform antiferromagnetic (AFM), helical and skyrmion crystal (SkX) phases. The averaged skyrmion density $\rho_{\textrm{sky}}$ is measured in arbitrary units. The film effectively has an easy axis (plane) for $\mathcal{K}<0$ ($\mathcal{K}>0$). The insets show the spatial configuration of the N\'{e}el order (yellow arrow) within the AFM phase.}
\label{Fig2}
\end{center}
\end{figure}

{\it Phase diagram.}|Interplay between exchange, aniso\- tropy and DM interactions allows for the stabilization of individual AFM skyrmions. This is not the case for the SkX phase, which also requires the polarizing effect of the staggered field on the N\'{e}el order, i.e., the presence of a Zeeman field. Fig. \ref{Fig2} illustrates the phase diagram of the AFM thin film at zero temperature in the parameter space $(\mathcal{K},\mathscr{B}_{\textrm{stag}})$, which contains the (uniform) AFM, helical and SkX phases. It has been obtained by computing the free-energy density \eqref{eq1} for each phase with a one-dimensional circularly-symmetric variational ansatz, along the lines of Ref. \onlinecite{Banerjee-PRX2014}. The SkX phase is found to be a ground state in a substantial area of the phase diagram: For $0\leq\mathscr{B}<\mathscr{B}_{F}$ (effective easy-axis anisotropy), the functional \eqref{eq1} coincides with the free energy of a ferromagnetic film with the broken reflection symmetry normal to the basal plane \cite{Bogdanov-JMMM1994}; as a result, the two first-order phase transitions helical $\Rightarrow$ SkX $\Rightarrow$ AFM are fomented in our AFM film at low temperatures when the staggered field is swept. This same behavior is also found in the spin-flop region ($\mathscr{B}>\mathscr{B}_{F}$), where the SkX phase is more robust and occupies a wider area. 

One consequence of Eq. \eqref{eq2} is that sublattice and time-reversal symmetries must break down to generate a net skyrmion charge in the AFM film \cite{Keesman-PRB2016}. This symmetry breaking occurs when $\mathcal{B}_{\textrm{stag}}$ is present, since the latter acts as a Zeeman field that establishes a preferred polarization (along the $z$ axis) for the order parameter \cite{FN1}. Skyrmions are magnetic excitations arising on top of the ensuing uniform AFM state and, since their polarization is fixed far from the core, so is their topological charge. Furthermore, the spin-transfer effect by an injected current on the skyrmion background also favors energetically skyrmions with a definite charge \cite{Ochoa-PRB2016}.

{\it Skyrmion transport.}|Skyrmions in AFM thin films are well-known to exhibit no gyrotropic response (Magnus-like force) \cite{Kim-PRB2015,Baker-PRL2016,Velkov-NJP2016}, which yields the suppression of the skyrmion Hall effect in these platforms, unlike their ferromagnetic counterparts \cite{Kong-PRL2013,Schutte-PRB2014,Jiang-NPhys2017}. In the gas phase, consisting of a dilute concentration of skyrmions embedded in a uniform AFM background, current-driven dynamics of these solitons are described by the terminal center-of-mass velocity $\partial_{t}\bm{X}|_{\textrm{t}}=-\zeta^{\textrm{gas}}_{\parallel}\,\bm{j}-\zeta^{\textrm{gas}}_{\perp}\,\hat{e}_{z}\times\bm{j}$, with $\zeta^{\textrm{gas}}_{\parallel}=\vartheta_{1}/\alpha s$ and $\zeta^{\textrm{gas}}_{\perp}=4\pi\vartheta_{2}\mathcal{Q}_{\textrm{sky}}/\alpha s\mathcal{I}$. Here, $\alpha$ denotes the Gilbert damping constant, $\mathcal{I}$ is a dimensionless geometric factor determined by the skyrmion texture, and the phenomenological constants $\vartheta_{1}$ and $\vartheta_{2}$, which parametrize the dissipative and reactive parts of the spin torque, respectively, depend on the interplay of spin-orbit and spin-transfer physics at the interface \cite{SM,FN2}. 

\begin{figure}[t]
\begin{center}
\includegraphics[width=8.75cm]{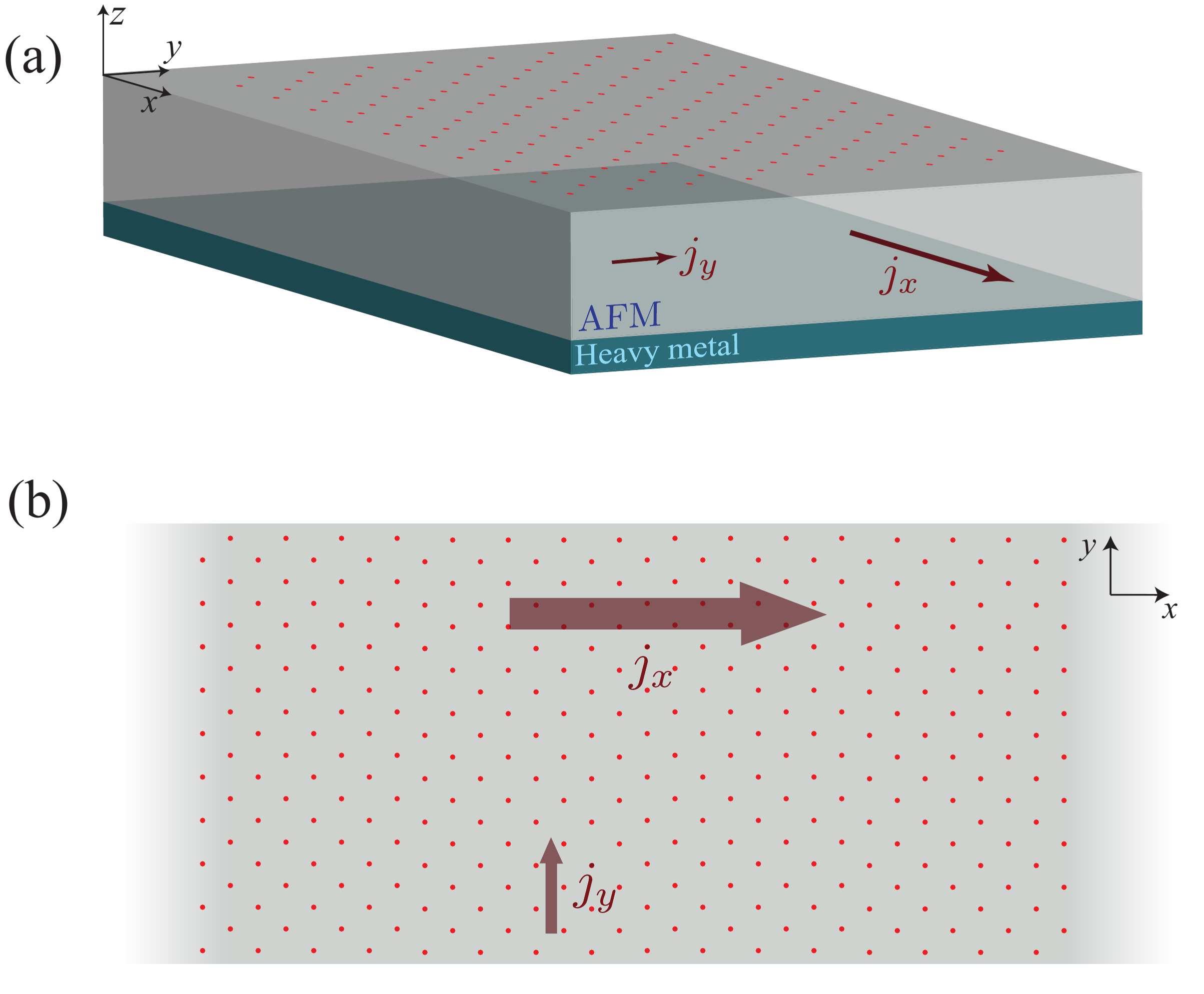}
\caption{(Infinite) Slab geometry for the transport of skyrmions in the crystal phase: (a) perspective and (b) top views. Red dots depict skyrmions located at the sites of the static hexagonal lattice. The slab extends along the $x$ direction (represented by the faded vertical boundaries), whose length is much larger than the separation between the contacts through which the electric current $j_{y}$ is injected.}
\label{Fig3}
\end{center}
\end{figure}

Let $\bm{u}=(u_{x}[\bm{r},t],u_{y}[\bm{r},t])$ be the collective coordinates parametrizing, in the continuum limit, the displacement of skyrmions in the crystal with respect to their equilibrium positions. For an infinite geometry along the $x$ direction (so that we can disregard boundary effects and assume translational invariance), see Fig. \ref{Fig3}, the motion of the skyrmion crystal in the steady state is described by the terminal velocity $\partial_{t}u_{x}|_{\textrm{t}}=-\zeta_{\parallel}^{\textrm{gas}}j_{x}+\zeta_{\perp}^{\textrm{gas}}j_{y}$ and $\partial_{t}u_{y}|_{\textrm{t}}=-p_{\textrm{gas}}^{\textrm{SkX}}(\zeta_{\perp}^{\textrm{gas}}j_{x}+\zeta_{\parallel}^{\textrm{gas}}j_{y})$. The reduction of its $y$-component as compared to the single-skyrmion counterpart, which is quantified by the dimensionless prefactor $p_{\textrm{gas}}^{\textrm{SkX}}<1$, originates in the loss of angular momentum via spin pumping into the metallic terminals \cite{SM}.

{\it Spin-Hall transport.}|In the adiabatic limit for spin dynamics, itinerant electrons experience a fictitious electromagnetic field when moving within the skyrmion-crystal \cite{Cheng-PRB2012}. For simplicity, we consider the case in which electrons keep their spins parallel or antiparallel to the N\'{e}el-order background without any deviation~\cite{FN3}. This leads to the following expressions for the emergent fields for electrons:
\begin{align}
\label{eq7}
\mathscr{B}_{z}&=\mp\frac{\hbar}{2e}\,\bm{l}\cdot(\partial_{x}\bm{l}\times\partial_{y}\bm{l})=\pm\frac{h}{e}\,\rho_{\textrm{sky}},\\
\bm{\mathcal{E}}&=\mp\frac{\hbar}{2e}\,\bm{l}\cdot(\nabla\bm{l}\times\partial_{t}\bm{l})=\mathscr{B}_{z}\,\hat{e}_{z}\times\partial_{t}\bm{u},
\end{align}
where $e$ is the electron charge and the sign $\pm$ corresponds to the spin-up(-down) bands with respect to $\bm{l}$. Here, we have disregarded other terms originating in Rasbha (spin-orbit) physics and spin-flip processes \cite{Yamane-PRBRC2016}. As a result, electrons flowing through the AFM SkX will exhibit a spin-Hall response, which can be understood as two copies of the topological Hall effect, one for the spin-up band and the other for the spin-down band. Within a semiclassical framework for the spin-Hall effect based on the Drude model \cite{Chudnovsky-PRL2007}, the SkX phase engenders a spin-Hall current:
\begin{align}
\label{eq9}
\bm{j}^{s,\textrm{SkX}}_{z}&=\pi(\hbar/e)^{2}\mu\hspace{0.05cm}\rho_{\textrm{sky}}\hspace{0.05cm}\hat{e}_{z}\times\big(ne\,\partial_{t}\bm{u}|_{\textrm{t}}-\bm{j}\big),
\end{align}
where the polarization direction of the spin current is along the local N\'{e}el order $\bm{l}$. Here, $n$, $\mu=\tau e/m_{\star}$ and $m_{\star}$ denote the concentration, mobility and effective mass of conduction electrons, respectively, and $\tau$ represents the scattering time \cite{SM}. The resultant spin current flowing through the interface of the AFM film with a proximate heavy metal can be measured by using the inverse spin Hall effect in the latter, and it is given by $\langle \bm{l} \cdot \hat{e}_{z} \rangle \bm{j}^{s,\textrm{SkX}}_{z}$ with the polarization along the global $z$ axis, where the prefactor $\langle \bm{l} \cdot \hat{e}_{z} \rangle$ is the spatially averaged (over the interface) projection of $\bm{l}$ onto the $z$ axis. This expression has been derived under the assumption that the skyrmion density is smooth on the scale of the electron scattering length. 

{\it Experimental platforms.}|A first realization of the fieldlike-torque scenario is provided by thin films made of chromia ($\alpha$-Cr$_{2}$O$_{3}$), in which an unconventional surface magnetism coexists with the bulk N\'{e}el phase below the N\'{e}el temperature \cite{Binek-NatMat2010}. Strikingly, this surface magnetization, rooted in the magnetoelectric effect, is collinear with the N\'{e}el order. As a result, the AFM order parameter couples linearly to the external magnetic field through the boundaries, yielding an effective Zeeman torque for the spin dynamics of chromia \cite{Zarzuela-PRB2018}. A second realization consists of AFM thin films subjected to spin exchange with an adjacent hard ferromagnet. The ensuing exchange bias effect \cite{Park-NMat2011,Wu-NPhys2011} allows for the control of the N\'{e}el order, since the magnetization of the ferromagnet (taken to be fully spin-polarized by, e.g., an external magnetic field) plays the role of the Zeeman field in our approach. This effect is also interfacial in nature and will be enhanced for uncompensated surfaces.

Metallic bipartite antiferromagnets such as Mn$_{2}$Au and CuMnAs offer a third platform for the stabilization of the SkX phase. These centrosymmetric materials, whose magnetic sublattices $\{A,B\}$ form inversion partners but, individually, have broken inversion symmetry, exhibit the recently discovered Edelstein spin-orbit torque \cite{Zeleny-PRL2014,Wadley-Sci2016}: Effective magnetic fields $\mathcal{B}_{A}=-\mathcal{B}_{B}\propto\hat{e}\times\bm{j}$ are induced on each sublattice via the inverse spin-galvanic (Edelstein) effect, where $\hat{e}$ denotes the crystallographic direction along which sublattice inversion symmetry is broken and where sign alternation stems from the aforementioned inversion partnership. Therefore, the nonequilibrium torque exerted by the charge current on the total spin density $s\bm{m}=\bm{M}_{A}+\bm{M}_{B}$ (with $s$ being the saturated spin density) becomes $s\partial_{t} \bm{m}|^{\textrm{neq}}_{\textrm{SO}}=\bm{l}\times\mathcal{B}_{\textrm{stag}}$, where the staggered field is defined per $2\mathcal{B}_{\textrm{stag}}=\mathcal{B}_{A}-\mathcal{B}_{B}$ \cite{Gomonay-PRL2016}. 

Even though we are invoking here a nonequilibrium effect, the stability of the AFM phases depicted in Fig. \ref{Fig2} applies to the low-current regime where skyrmions are pinned by defects. CuMnAs appear to exhibit uniaxial MCA for thin enough films \cite{Wadley-SciRep2015}. Note that the axes of the phase diagram are the staggered field and the effective uniaxial anisotropy, which we can control by changing the charge current and the external field, respectively. This real-time controllability allows us to explore an entire phase diagram with one sample. Furthermore, choice of the $x$ axis along $\hat{e}$, see Fig. \ref{Fig3}, makes the charge current play two roles: $j_{y}$, which is fixed in our setup, allows for the stabilization of the SkX phase via the staggered field, whereas the other independent (and tunable) component of the charge current, $j_{x}$, is employed to control the drag of the skyrmion lattice.

{\it Discussion.}|Spin currents offer a knob to inject and drive AFM skyrmions within the insulating medium, akin to the ferromagnetic case \cite{Ochoa-PRB2016,Ochoa-PRBRC2017}. In that regard, a two-terminal geometry allows for the pumping of topological charge into the antiferromagnet via the spin-transfer effect; measurements of the ensuing (long-range) spin drag signal could be used to i) probe the existence of these topological textures, and ii) discriminate the gas and SkX phases, since the drag coefficient exhibits a different dependence on the staggered field \cite{Ochoa-PRBRC2017}. Other experimental techniques well suited to read out the SkX phase could be the x-ray magnetic linear dichroism (XMLD) imaging, the spin-transfer-torque ferromagnetic resonance, and the noncolinear magnetoresistance \cite{Hanneken-2017}, to name a few. 
 
As can be inferred from Fig. \ref{Fig2}, the presence of DM interactions at the bulk level enhances the stabilization of the SkX phase. Furthermore, reactive spin-transfer torques yield gyrotropic terms in the equations of motion for the skyrmion crystal, which therefore exhibits current-driven dynamics for the steady state analogous to that of the ferromagnetic case \cite{Ochoa-PRBRC2017}. It is worth remarking here that an additional anomalous contribution to the spin Hall current arises from the nontrivial Berry curvature of the Bloch bands, see for instance Ref. \onlinecite{Gobel-PRBRC2017} for a numerical study on hexagonal AFM lattices. 

Finally, compared to other recent proposals, for which the SkX phase emerges in synthetic AFM trilayers \cite{Buhl-PSSRRL2017} or in a ferromagnetic honeycomb lattice exchange-coupled to an AFM substrate \cite{Gobel-PRBRC2017}, our approach to the stabilization of the SkX phase can be applied to a wider class of natural (intrinsic) antiferromagnets and does not rely on a fine engineering of the AFM heterostructures. One relevant question to be elucidated in future research is the effect of disorder on the skyrmion crystal. For weak disorder we would expect a glassy behavior (as in, e.g., type-II superconductors in the Abrikosov state). Furthermore, since the dimensionality of our system is two, the emergence of a Bragg glass phase is feasible too \cite{Giamarchi-PRB1995,Giamarchi-PhysC2000}. We also expect the low-energy excitation spectrum of the disordered SkX phase to be gapped, as in the ferromagnetic counterpart \cite{Hoshino-PRB2017}.

\section*{Acknowledgments}
We thank J. Sinova for insightful discussions on the Mn$_{2}$Au and CuMnAs platforms. This work has been supported by NSF under Grant No. DMR-1742928. R.Z. thanks Fundaci\'{o}n Ram\'{o}n Areces for support through a postdoctoral fellowship within the XXVII Convocatoria de Becas para Ampliaci\'{o}n de Estudios en el Extranjero en Ciencias de la Vida y de la Materia.

{\it Note added}: Recently, we noticed that our theory for the fieldlike-torque-induced realization of AFM skyrmion crystals can explain a recent observation of antiferromagnetic skyrmions reported in Ref. \cite{Wang-Nano2018}, which corresponds to the AFM film/ferromagnet platform proposed in this Letter.

\onecolumngrid
 
 \newpage
 
\hspace{0.5cm}
\section*{\Large Supplemental Material}
\hspace{0.5cm}

\section*{1. Lagrangian formalism and dynamics of bipartite antiferromagnets}
We regard thin films as quasi-two-dimensional (quasi-2D) systems along the $xy$ plane, which we take to be isotropic at the coarse-grained level. An effective low-energy theory for bipartite antiferromagnets can be developed in terms of two continuum coarse-grained fields, namely the (staggered) N\'{e}el order $\bm{l}$ and the spin density $s\hspace{0.03cm}\bm{m}$ \cite{AFM}. Here, $s=\hbar S d/\mathcal{V}$ is the saturated spin density, where $d$ denotes the film thickness and $\hbar S$, $\mathcal{V}$ are the spin and volume per site, respectively. These fields satisfy the nonlinear local constraints $\bm{l}^{2}=1$ and $\bm{l}\cdot\bm{m}=0$, and the presence of a well-developed N\'{e}el order implies $|\bm{m}|\ll1$ on the scale of the exchange coupling. To the lowest order in these coarse-grained fields, the quasi-2D Lagrangian density in the continuum limit becomes
\begin{equation}
\label{SM1}
\mathcal{L}_{\textrm{AFM}}[t;\bm{l},\bm{m}]=s\,\bm{m}\cdot(\bm{l}\times\partial_{t}\bm{l})-\mathcal{F}_{\textrm{AFM}}[\bm{l},\bm{m}],\hspace{0.5cm}\mathcal{F}_{\textrm{AFM}}[\bm{l},\bm{m}]=\frac{\bm{m}^{2}}{2\chi}-\bm{m}\cdot\mathcal{B}_{\textrm{ext}}+\mathcal{F}_{\textrm{stag}}[\bm{l}],
\end{equation}
where $\chi$ denotes the (transverse) spin susceptibility and $\mathcal{B}_{\textrm{ext}}=\gamma s\bm{B}_{\textrm{ext}}$ represents the normalized (external) magnetic field. The first term describes the kinetic (Berry-phase) Lagrangian and $\mathcal{F}_{\textrm{stag}}$ represents the minimal model for the total energy for the staggered order parameter, which contains isotropic exchange and uniaxial anisotropy contributions:
\begin{equation}
\label{SM2}
\mathcal{F}_{\textrm{stag}}[\bm{l}]=\frac{A}{2}\sum_{\mu=x,y}(\partial_{\mu}\bm{l})^{2}+\frac{K}{2} l_{z}^{2},
\end{equation}
where $A$ and $K$ are the stiffness and anisotropy constants, respectively. Both $A$ and $\chi^{-1}$ are proportional to $JS^{2}$, with $J$ being the microscopic exchange energy, and $K<0$ describes the hard (anisotropy) $xy$ plane. Spin dynamics of AFM films are ruled by the following Landau-Lifshitz-Gilbert-type equations \cite{Zarzuela-PRBRC2017}:
\begin{align}
s\partial_{t}\bm{l}&=\bm{l}\times\bm{f}_{m}, \label{SM3}\\
s\partial_{t}\bm{m}&=\bm{l}\times[\bm{f}_{l}-s\alpha\partial_{t}\bm{l}]+\bm{m}\times\bm{f}_{m}+\bm{\tau}_{m}, \label{SM4}
\end{align}
where $\bm{f}_{l}=-\delta_{\bm{l}}F_{\textrm{AFM}}$ and $\bm{f}_{m}=-\delta_{\bm{m}}F_\textrm{AFM}$ are the thermodynamic forces conjugate to the N\'{e}el order and the spin density, respectively. Here, $\alpha$ denotes the Gilbert-damping constant, where we have considered an isotropic Gilbert-Rayleigh dissipation function $\mathcal{R}[\bm{l}]=s\alpha(\partial_{t}\bm{l})^{2}/2$ for simplicity. 

The magnetic torque $\bm{\tau}_{m}$ has two contributions: The first accounts for spin-transfer processes between the localized spins of the AFM lattice and itinerant electrons. We restrict our considerations to the texture-induced spin-transfer torque since spin-orbit torques do not exert an effective force on the skyrmion texture for the rigid ansatz \cite{Lin-PRB2017}. It reads as $\bm{\tau}_{m,\textrm{ST}}=\vartheta_{1}\bm{l}\times(\bm{j}\cdot\nabla)\bm{l}+\vartheta_{2}(\bm{j}\cdot\nabla)\bm{l}$, where $\bm{j}$ is the charge current and $\vartheta_{1},\vartheta_{2}$ are phenomenological parameters depending on the interplay of spin-orbit and spin-transfer physics at the interface. The second contribution consists of the fieldlike torque $\bm{\tau}_{m,\textrm{FL}}=\bm{l}\times\mathcal{B}_{\textrm{stag}}$, with $\mathcal{B}_{\textrm{stag}}$ denoting the staggered field. The latter can be absorbed into the reactive torque of Eq. \eqref{SM4} via the redefinition $\bm{f}_{l}\rightarrow \bm{f}_{l}+\mathcal{B}_{\textrm{stag}}$ of the thermodynamic force, which is equivalent to adding the Zeeman-like coupling $-\bm{l}\cdot\mathcal{B}_{\textrm{stag}}$ to the energy $\mathcal{F}_{\textrm{stag}}$.

An effective Lagrangian for the N\'{e}el order can be obtained by integrating out the spin field $\bm{m}$ subject to the local constraints $\bm{l}^{2}=1$ and $\bm{l}\cdot\bm{m}=0$. The resultant constitutive relation reads as $\bm{m}=\chi\,\bm{l}\times\left[s\partial_{t}\bm{l}+\mathcal{B}_{\textrm{ext}}\times\bm{l}\right]$ and the effective Lagrangian becomes:
\begin{equation}
\label{SM5}
\mathcal{L}_{\textrm{eff}}[t;\bm{l}]=\frac{\chi s^{2}}{2}(\partial_{t}\bm{l})^{2}+\mathcal{L}_{\textrm{eff}}^{B}[\bm{l}]-\mathcal{F}_{\textrm{eff}}[\bm{l}],\hspace{0.5cm}\mathcal{L}_{\textrm{eff}}^{B}[\bm{l}]=\bm{a}[\bm{l}]\cdot\partial_{t}\bm{l},\hspace{0.5cm}\mathcal{F}_{\textrm{eff}}[\bm{l}]=\frac{\chi}{2}\left(\bm{l}\times\mathcal{B}_{\textrm{ext}}\right)^{2}+\mathcal{F}_{\textrm{stag}}[\bm{l}],
\end{equation}
where the first and second terms account for the inertia and the geometric Berry phase for the dynamics of the N\'{e}el order, respectively, with $\bm{a}[\bm{l}]=\chi s (\mathcal{B}_{\textrm{ext}}\times\bm{l})$. Note that $\mathcal{F}_{\textrm{eff}}[\bm{l}]$ reduces to Eq. (1) in the main text when both magnetic and staggered field are aligned along the $z$ axis (in the absence of DM interactions).

\section*{2. Skyrmion dynamics in the collective-variable approach}

\subsection*{Gas phase}

Within the collective-variable approach, the time dependence of the N\'{e}el order is encoded in the skyrmion center of mass, $\bm{l}(t,\bm{r})\equiv\bm{l}[\bm{X}(t)]=\bm{l}[\bm{r}-\bm{X}(t)]$, whose coordinates describe the soft modes of the texture (for rigid skyrmions). Therefore $\partial_{t}\bm{l}=\nabla_{\dot{\bm{X}}(t)}\bm{l}=-(\dot{\bm{X}}\cdot\nabla_{\bm{r}})\bm{l}$ denotes the covariant derivative of the N\'{e}el order along the trajectory of the skyrmion core. The conjugate momentum to $\bm{X}$, $\bm{\Pi}[\bm{r}-\bm{X},\dot{\bm{X}}]=\delta_{\dot{\bm{X}}} L_{\textrm{eff}}$, defines a trajectory within a section of the tangent bundle $T\mathbb{R}^{2}$. As a result, the total derivative $D_{t}\bm{\Pi}$ must be interpreted as a covariant derivative along the curve tangent to the center of mass:
\begin{equation}
\label{SM6}
D_{t}\bm{\Pi}=-\dot{X}_{k}\partial_{k}\bm{\Pi}+\ddot{X}_{k}\partial_{v_{k}}\bm{\Pi}=\hat{e}_{i}(M_{ij}[\bm{l}]+M^{B}_{ij}[\bm{l}])\ddot{X}_{j}+\hat{e}_{i}G^{B}_{ij}[\bm{l}]\dot{X}_{j}+\hat{e}_{i}\mathcal{R}_{ijk}[\bm{l}]\dot{X}_{j}\dot{X}_{k}+\hat{e}_{i}\widetilde{\mathcal{R}}_{ijk}[\bm{l}]\dot{X}_{j}\ddot{X}_{k}.
\end{equation}
The tensors $[M]$, $[M^{B}]$, 
$[\mathcal{R}]$ and $[\widetilde{\mathcal{R}}]$,  are defined per:
\begin{align}
\label{SM7}
M_{ij}[\bm{l}]&=\chi s^{2}\int_{S}\ud^{2}\bm{r}\hspace{0.15cm}\partial_{i}\bm{l}\cdot\partial_{j}\bm{l},\\
M_{ij}^{B}[\bm{l}]&=-\chi s\epsilon^{\alpha\beta\gamma}\mathscr{B}_{\textrm{ext},\alpha}\int_{S}\ud^{2}\bm{r}\hspace{0.15cm}\partial_{v_{j}}\left[l_{\beta}\partial_{i}l_{\gamma}\right],\\
\mathcal{R}_{ijk}[\bm{l}]&=-\chi s^{2}\int_{S}\ud^{2}\bm{r}\hspace{0.15cm}\partial_{k}\left[\partial_{i}\bm{l}\cdot\partial_{j}\bm{l}\right],\\
\widetilde{\mathcal{R}}_{ijk}[\bm{l}]&=\chi s^{2}\int_{S}\ud^{2}\bm{r}\hspace{0.15cm}\partial_{v_{k}}\left[\partial_{i}\bm{l}\cdot\partial_{j}\bm{l}\right],
\end{align}
where $S$ is the surface of the AFM film and $\epsilon^{\alpha\beta\gamma}$ denotes the components of the Levi-Civita tensor. The boundary condition $\bm{l}|_{\partial S}=\hat{e}_{z}$ on the N\'{e}el texture makes the tensor 
$[\mathcal{R}]$ be identically zero. Furthermore, the inertia tensor reduces to a scalar for the hard cut-off variational ansatz: The effective inertia reads $M=\chi s^{2}\mathcal{I}$, where $\mathcal{I}=(\pi^{2}+\gamma-\textrm{Ci}(2\pi)+\ln2\pi)\pi/2$ is a dimensionless geometric factor. Here, $\gamma\sim0.57721$ is the Euler-Mascheroni constant and Ci($z$) denotes the cosine integral function. Note that the tensors $[M^{B}]$ and $[\widetilde{\mathcal{R}}]$ capture the effect of the (inertial) deformation of the skyrmion on the motion of the collective variables. However, if we restrict ourselves to a (slow) dynamic regime where the rigidity of the soliton is preserved, the components of these two tensors vanish. 

Another contribution to the equation of motion for single skyrmions stems from the term $\delta_{\bm{X}}L_{\textrm{eff}}^{B}$, which, for $\mathcal{B}_{\textrm{ext}}\propto\hat{e}_{z}$, takes the form 
$\hat{e}_{i}\widetilde{G}^{B}_{ij}[\bm{l}]\dot{X}_{j}$. The total gyrotropic tensor $[G^{B}]+[\widetilde{G}^{B}]$ reads as 
\begin{align}
\label{SM8}
G_{ij}^{B}[\bm{l}]+\widetilde{G}_{ij}^{B}[\bm{l}]&=2\chi s\epsilon^{ijz}\mathscr{B}_{\textrm{ext}}\int_{S}\ud^{2}\bm{r}\hspace{0.15cm}(\partial_{x}\bm{l}\times\partial_{y}\bm{l})\cdot\hat{e}_{z}=-\chi s\epsilon^{ijz}\mathscr{B}_{\textrm{ext}}\oint_{\partial S}\ud\bm{r}\cdot\left[(\nabla\bm{l}\times\bm{l})\cdot\hat{e}_{z}\right].
\end{align}
The second identity is derived from the Green-Stokes theorem and the line integral is performed over the surface boundary. Again, the boundary condition $\bm{l}|_{\partial S}=\hat{e}_{z}$
yields the vanishing of the integrand and, therefore, to the absence of a magnetic-field-induced gyrotropic response for AFM skyrmions. Dissipation is modeled by the Gilbert-Rayleigh function $\mathcal{R}[\bm{l}]=s\alpha(\partial_{t}\bm{l})^{2}/2$ \cite{Gilbert-1955}, which provides the dominant term in the low-frequency (compared to the microscopic exchange $J$) regime. Within the collective-variable approach it becomes
\begin{equation}
\label{SM9}
\mathcal{R}[\bm{l}]=\frac{1}{2}\dot{\bm{X}}\cdot[\Gamma]\cdot\dot{\bm{X}},\hspace{0.5cm}\Gamma_{ij}=\frac{M_{ij}}{\mathcal{T}},
\end{equation}
where $\mathcal{T}=\chi s/\alpha$ represents a relaxation time. In summary, the equation of motion for single skyrmions (so-called Thiele equation) turns out to be:
\begin{equation}
\label{SM10}
M\ddot{\bm{X}}+\frac{M}{\mathcal{T}}\dot{\bm{X}}=\bm{F}+\bm{F}_{J}
\end{equation}
where $\bm{F}\equiv-\delta_{\bm{X}}{F}_{\textrm{eff}}$ is the conservative force and $\bm{F}_{J}$ is the external force corresponding to the magnetic torque $\bm{\tau}_{m,\textrm{ST}}$. The latter reads as $\bm{F}_{J}=-\vartheta_{1}\mathcal{I}\,\bm{j}-4\pi\vartheta_{2}\mathcal{Q}_{\textrm{sky}}\hat{e}_{z}\times\bm{j}$  \cite{Zarzuela-PRB2018}.

\begin{figure}[t]
\begin{center}
\includegraphics[width=0.8\textwidth]{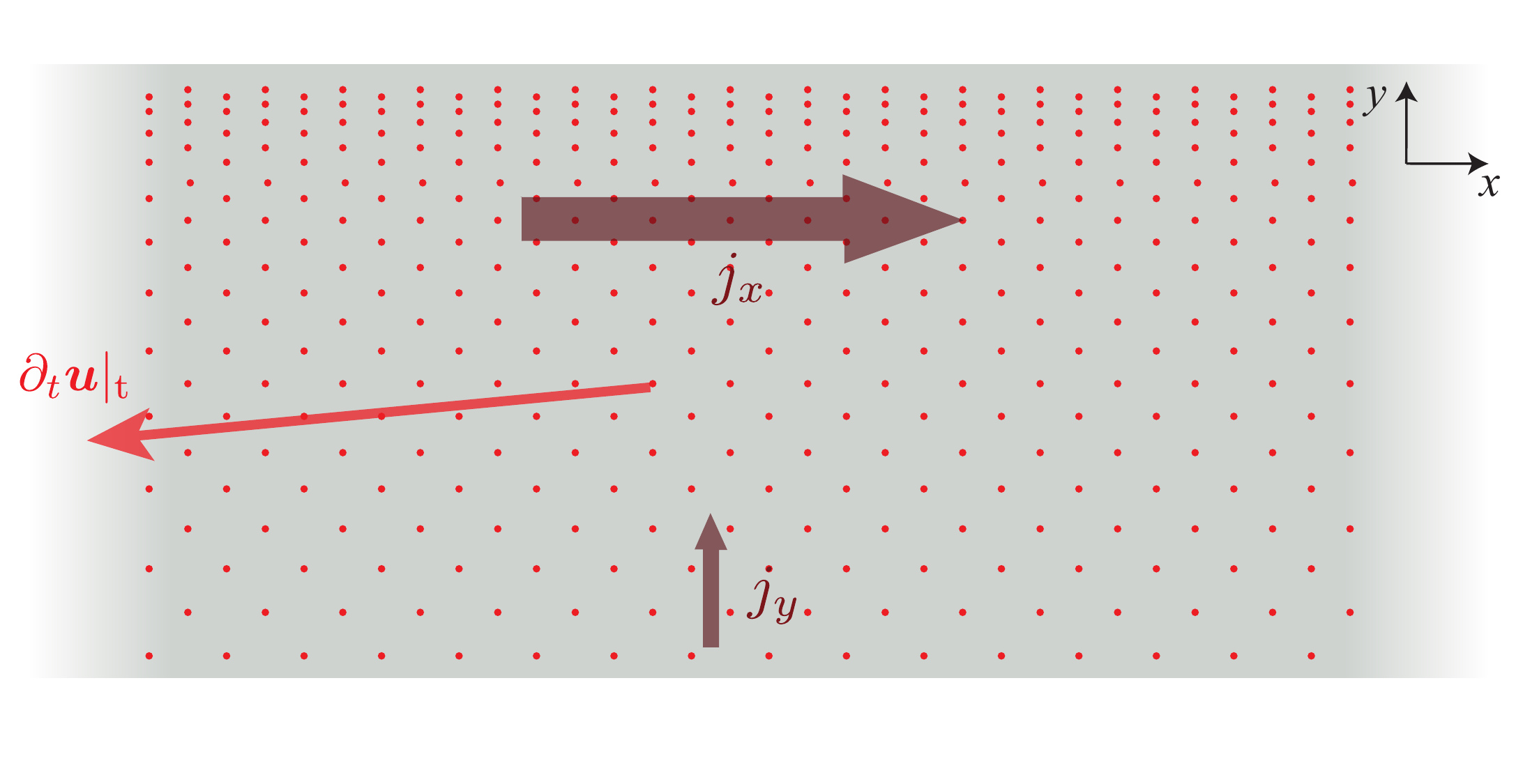}
\caption{Top view of the (infinite) slab geometry for the transport of skyrmions in the crystal phase. Red dots depict skyrmions located at the sites of the hexagonal crystal lattice. In the steady state, the skyrmion lattice moves along the direction represented by the red arrow. The slab extends along the $x$ direction (represented by the faded vertical boundaries), whose length is much larger than the separation between the metallic contacts through which the electric current $j_{y}$ is injected.}
\label{Fig4}
\end{center}
\end{figure}

\subsection*{Crystal phase} 

The crystal phase is described, within the collective-variable approach, by the displacement fields $u_{x,y}(\bm{r},t)$ of skyrmions with respect to their equilibrium (lattice) positions in the $xy$ plane. Once more, dynamics of the N\'{e}el order are encoded in the time evolution of these fields, $\bm{l}[\bm{r},t]=\bm{l}[\bm{r}-\bm{u}(\bm{r},t)]$, where $\bm{u}$ is smooth over length scales of the order of the lattice constant. By proceeding along the lines of the previous section (except that integrals are now restricted to the unit cell, with one skyrmion located at its center), we obtain the following quasi-2D Lagrangian density for the crystal in the continuum limit \cite{Petrova-PRB2011}:
\begin{equation}
\label{SM11}
\mathcal{L}_{\textrm{SkX}}[t,\bm{u}]=\frac{\rho_{M}}{2}\left[\dot{u}_{x}^{2}+\dot{u}_{y}^{2}\right]-\frac{1}{2}\sum_{i,j=x,y}\left[\lambda (u_{ii})^{2}+2\mu\,u_{ij}u_{ij}\right],
\end{equation}
where the first term accounts for the inertia of the dynamics of the displacement fields and the second term describes the low-energy elastic excitations of the skyrmion crystal. Here, $\rho_{M}=M/\mathcal{S}$ is the inertia density (with $\mathcal{S}$ being the area of the unit cell), $u_{ij}=(\partial_{i}u_{j}+\partial_{j}u_{i})/2$ denotes the components of the strain tensor and $(\lambda,\mu)/d\sim D^{2}/A$ represent the Lam\'{e} coefficients, which characterize the strain-stress relationship in elastic media with trigonal/hexagonal symmetry. In the above expression we have disregarded any possible bending along the $z$ axis. The Euler-Lagrange equations of motion become
\begin{equation}
\label{SM12}
\rho_{M}\ddot{u}_{i}+\frac{\rho_{M}}{\mathcal{T}}\dot{u}_{i}=\lambda\,\partial_{i}u_{kk}+2\mu\,\partial_{k}u_{ik}+F_{J,i}, \hspace{0.5cm}i,k=x,y,
\end{equation}
where we have again accounted for dissipative processes via the Gilbert-Rayleigh dissipation function and now $\bm{F}_{J}$ represents the force (surface) density associated with $\bm{\tau}_{m,\textrm{ST}}$. For the infinite slab geometry depicted in Fig. \ref{Fig4}, we assume the thin film to be translational invariant along the $x$ axis, i.e., $u_{x,y}\equiv u_{x,y}(y)$. Furthermore, boundary conditions (BCs) are imposed at the interfaces $y=0$ and $y=L$ through the balance between the applied tension and the internal stress, $\bm{F}_{\textrm{pump}}=\hat{e}_{i}\sigma_{ij}n_{j}$, where $\bm{n}$ denotes the normal to the interface, $\sigma_{ij}=\lambda u_{kk}\delta_{ij}+2\mu\,u_{ij}$ are the components of the stress tensor, and the tension $\bm{F}_{\textrm{pump}}$ quantifies the dissipation (pumping) of angular momentum into the terminals induced by the dynamics of the skyrmion crystal. The expression for the latter reads as \cite{Ochoa-PRBRC2017}:
\begin{equation}
\label{SM13}
\bm{F}_{\textrm{pump}}=-\left(\frac{h\rho_{\textrm{sky}}}{e}\right)^{2}g\hspace{0.03cm}\dot{\bm{u}}|_{\textrm{b}},
\end{equation}
where $\rho_{\textrm{sky}}\equiv\mathcal{Q}_{\textrm{sky}}/\mathcal{S}$ is the skyrmion charge density averaged over the unit cell, $g$ is the effective interfacial conductance and $\dot{\bm{u}}|_{\textrm{b}}$ denotes the velocity of the skyrmion crystal perpendicular to the boundary. We obtain the following system of differential equations in the steady state:
\begin{equation}
\label{SM14}
\left\{
\begin{array}{ll}
\frac{\rho_{M}}{\mathcal{T}}\dot{u}_{x}|_{\textrm{t}}& = \mu\partial_{y}^{2}u_{x}+F_{J,x},\\
\frac{\rho_{M}}{\mathcal{T}}\dot{u}_{y}|_{\textrm{t}}&= (\lambda+2\mu)\partial_{y}^{2}u_{y}+F_{J,y},\end{array}\right.\hspace{0.8cm}\textrm{BCs}:
\left\{\begin{array}{ll}
\partial_{y}u_{x}(0)=\partial_{y}u_{x}(L)=0,\\
-\partial_{y}u_{y}(0)=\partial_{y}u_{y}(L)=\left(\frac{h\rho_{\textrm{sky}}}{e}\right)^{2}\frac{g\hspace{0.03cm}\dot{u}_{y}|_{\textrm{t}}}{\lambda+2\mu},
\end{array}\right.
\end{equation}
whose solution for the displacement fields $\bm{u}$ yields the following expression for the terminal velocity of the skyrmion crystal:
\begin{equation}
\label{SM15}
\dot{u}_{x}|_{\textrm{t}}=-\zeta_{\parallel}^{\textrm{gas}}j_{x}+\zeta_{\perp}^{\textrm{gas}}j_{y},\hspace{1.0cm}\dot{u}_{y}|_{\textrm{t}}=-p_{\textrm{gas}}^{\textrm{SkX}}\big(\zeta_{\perp}^{\textrm{gas}}j_{x}+\zeta_{\parallel}^{\textrm{gas}}j_{y}\big),
\end{equation}
with the dimensionless prefactor being 
\begin{equation}
\label{SM16}
p_{\textrm{gas}}^{\textrm{SkX}}=\left[1+2\left(\frac{h\rho_{\textrm{sky}}}{e}\right)^{2}\frac{g}{L}\frac{\mathcal{S}}{\alpha s\mathcal{I}}\right]^{-1}.
\end{equation}

\section*{3. Spin Hall transport}

Electrons flowing in the (metallic) AFM thin film are expected to exhibit a Hall response due to the presence of fictitious electromagnetic fields. The latter are (electron-)spin-dependent, see Eqs. (3) and (4) in the main text, leading to the generation of a spin current via the spin Hall effect. The corresponding spin-Hall current can be estimated along the lines of Ref. \onlinecite{Chudnovsky-PRL2007}, based on a Drude model for semiclassical transport: The Hamilton equations of motion for conduction electrons in the adiabatic limit read as
\begin{equation}
\label{SM17}
m_{\star}\ddot{\bm{r}}=-\frac{m_{\star}}{\tau}\dot{\bm{r}}+e\,\big[\bm{E}+\mathcal{E}(\bm{r},t,s_{z})\big]+e\,\dot{r}\times\mathcal{B}(\bm{r},t,s_{z}).
\end{equation}
Here, the external electric field $\bm{E}$ drives the injected current and we have considered the Drude model for collisions with a spin-independent scattering time $\tau$ and an effective mass $m_{\star}$. Since the emergent electromagnetic fields are proportional to $h/e$ [see Eqs. (3)-(4) in the main text], we can treat them as a perturbation in Eq. \eqref{SM17} and split the resultant solution into $\bm{r}=\bm{r}_{0}+\bm{r}_{1}$, where the dynamics of $\bm{r}_{0}$ are driven by the external fields and $\bm{r}_{1}$ represents a small spin-dependent contribution. In the steady state, we have the following identities:
\begin{align}
\label{SM18}
\langle\dot{\bm{r}}_{0}(t)\rangle&=\hat{\mu}\bm{E},\hspace{1.0cm}\hat{\mu}=\frac{e\tau}{m_{\star}}\mathbb{I}_{2},\\
\langle\dot{\bm{r}}_{1}(t,s_{z})\rangle&=\hat{\mu}\left[\langle\mathcal{E}(\bm{r},t,s_{z})\rangle+\langle\dot{\bm{r}}_{0}\rangle\times\langle\mathcal{B}(\bm{r},t,s_{z})\rangle\right],
\end{align}
where $\hat{\mu}$ is the mobility tensor and, once more, $\langle\cdots\rangle$ denotes average over the unit cell of the skyrmion crystal. The total (pseudo-)spin current is defined per $\vec{\bm{j}}^{s}=(\hbar/2)\llbracket\bm{\sigma}\otimes\langle\dot{\bm{r}}\rangle\rrbracket$, with $\llbracket\cdots\rrbracket\equiv\textrm{Tr}\big[\hat{\rho}\cdots\big]$ and $\hat{\rho}\in\textrm{SU}(2)$ being the density matrix operator for conduction electrons. Note that, in the rotated frame of reference, minority and majority bands (with respect to the N\'{e}el order) correspond to the spin-up and spin-down bands with respect to the quantization axis $\hat{e}_{z}$. Therefore, the concentration of carriers for the majority ($n_{+}$) and minority ($n_{-}$) species is given by $n_{\pm}=\textrm{Tr}[\hat{\rho}\,\hat{P}_{\mp}]$, where $\hat{P}_{\pm}=(\hat{\mathbb{I}}_{2}\pm\hat{\sigma}_{z})/2$ are the projection operators on the spin-up(-down) bands. The total concentration of carriers becomes $n=n_{+}+n_{-}$ and the operator $\hat{\rho}$ can be cast as
\begin{equation}
\label{SM19}
\hat{\rho}=\frac{n}{2}\left[\hat{\mathbb{I}}_{2}-p\,\hat{\sigma}_{z}\right],
\end{equation}
where $p=(n_{+}-n_{-})/n$ defines the net spin polarization of the electron fluid. With account of Eqs. \eqref{SM18}--\eqref{SM19} we obtain the following expression for the $z$-projection of the (pseudo-)spin current:
\begin{align}
\label{SM20}
\bm{j}^{s}_{z}&=(\hbar/2)\hspace{0.03cm}\hat{\mu}\big\llbracket\hat{\sigma}_{z}\big(\bm{E}+\langle\mathcal{E}(\bm{r},t,s_{z})\rangle\big)\big\rrbracket+(\hbar/2)\hspace{0.03cm}\hat{\mu}\Big[\hat{\mu}\bm{E}\times\big\llbracket\hat{\sigma}_{z}\langle\mathcal{B}(\bm{r},t,s_{z})\rangle\big\rrbracket\Big],\\
&=-(\hbar/2)\hspace{0.03cm}np\hspace{0.03cm}\hat{\mu}\bm{E}+\pi(\hbar/e)^{2}\rho_{\textrm{sky}}\hat{\mu}\Big[\hat{e}_{z}\times\big(ne\,\partial_{t}\bm{u}|_{\textrm{t}}-\bm{j}\big)\Big].\nonumber
\end{align}
where $\bm{j}=\sigma_{D}\bm{E}$ is the charge current and $\sigma_{D}=n\tau e^{2}/m_{\star}$ denotes the Drude conductivity. In the above expression we have assumed that the skyrmion texture (density) is smooth on the scale of the electron scattering length, i.e. $R_{\star}\ll v_{F}\tau$ with $v_{F}$ being the Fermi velocity. Note that the net spin polarization $p$ will be zero in the absence of an external magnetic field, so that the first term in Eq. \eqref{SM20} vanishes.

Finally, it is worth remarking that the polarization direction of the (pseudo-)spin current is along the local N{\'e}el order $\bm{l}$, which changes spatially. When we detect the physical spin current flowing through the interface of the AFM film with a heavy metal via the inverse spin Hall effect, the projection of the (pseudo-)spin current onto the global $z$ axis will be measured, resulting in the expression for the physical spin current (with the polarization direction along the $z$ axis), $\langle \bm{l} \cdot \hat{e}_z \rangle_{S} \bm{j}^s_z$, where $\langle \cdots \rangle_{S}$ represents the spatial average of the quantity over the interface.


\begin{thebibliography}{99}
\bibitem{FN1} Sublattice symmetry can also break down at the interface with the HM and/or at the structural level (defects, etc.).
\end{thebibliography}
\end{document}